\documentclass[review,times,nonatbib]{elsarticle} 

\usepackage{graphicx}
\usepackage{amsmath,amssymb}
\usepackage{algorithm}
\usepackage{algpseudocode}
\usepackage{subfig}
\usepackage{tabularx}
\usepackage{makecell}
\usepackage{xcolor}

\makeatletter
\let\c@author\relax
\makeatother

\usepackage[backend=biber,style=numeric-comp,sorting=none]{biblatex}
\renewcommand{\textcolor}[2]{#2}

\begin{document}

\begin{frontmatter}



\title{Leverage Knowledge Graph and Large Language Model for Law Article Recommendation: A Case Study of Chinese Criminal Law}


\author[label1]{Yongming Chen} 
\ead{cute-boy@sjtu.edu.cn}
\author[label2]{Miner Chen}
\ead{210730202@mail.dhu.edu.cn}
\author[label2]{Ye Zhu}
\ead{dhu_zhuye@163.com}
\author[label2]{Juan Pei}
\ead{3530372969@qq.com}
\author[label2]{Siyu Chen}
\ead{2212529533@qq.com}
\author[label2]{Yu Zhou}
\ead{2123157097@qq.com}
\author[label2]{Yi Wang}
\ead{Iris20416@126.com}
\author[label2]{Yifan Zhou}
\ead{2078613329@qq.com}
\author[label2]{Hao Li}
\ead{123470595@qq.com}
\author[label1]{Songan Zhang\corref{cor1}}
\ead{songanz@sjtu.edu.cn}
\cortext[cor1]{Corresponding author} 

\affiliation[label1]{organization={Global Institute of Future Technology, Shanghai Jiao Tong University},
            city={Shanghai},
            postcode={200240}, 
            state={Shanghai},
            country={China}}

\affiliation[label2]{organization={Donghua University}, 
            city={Shanghai},
            postcode={201620}, 
            state={Shanghai},
            country={China}}
            
\begin{abstract}
\textcolor{blue}{Judicial} efficiency is \textcolor{blue}{critical} to social stability. However, in many countries \textcolor{blue}{worldwide}, grassroots courts \textcolor{blue}{face substantial} case backlogs, and \textcolor{blue}{judicial} decisions \textcolor{blue}{remain} heavily dependent on judges' \textcolor{blue}{cognitive efforts}, with \textcolor{blue}{insufficient} intelligent tools to \textcolor{blue}{enhance} efficiency. To address this issue, we propose a \textcolor{blue}{highly efficient} law article recommendation approach \textcolor{blue}{combining} a Knowledge Graph (KG) \textcolor{blue}{and} a Large Language Model (LLM). First, we construct a Case-Enhanced Law Article Knowledge Graph (CLAKG) \textcolor{blue}{to} store current law articles, historical case information, and their \textcolor{blue}{interconnections}, \textcolor{blue}{alongside an LLM-based automated construction method}. \textcolor{blue}{Building on this}, we propose a closed-loop law article recommendation framework \textcolor{blue}{integrating graph embedding-based} retrieval \textcolor{blue}{and} KG-grounded LLM reasoning. Experiments on judgment documents from \textcolor{blue}{China Judgments Online} \textcolor{blue}{demonstrate} that our method \textcolor{blue}{boosts} law article recommendation accuracy from 0.549 to 0.694, \textcolor{blue}{outperforming strong baselines significantly}. \textcolor{blue}{To support reproducibility and future research, all source code and processed datasets are publicly available on GitHub (see Data Availability Statement)}.
\end{abstract}

\begin{highlights}
\item Proposed a novel Case-Enhanced Law Article Knowledge Graph (CLAKG) that integrates law articles and historical case data, enhancing the accuracy of law article recommendations.
\item Introduced an automated CLAKG construction method that uses a Large Language Model (LLM) to reduce manual input and improve scalability.
Developed a closed-loop recommendation system, leveraging LLMs and CLAKG for more accurate law article recommendations while mitigating common issues like hallucinations in LLM outputs.
\item Achieved significant improvement in accuracy: The proposed method improved law article recommendation accuracy from 0.549 to 0.694, outperforming baseline models such as BERT, DPCNN, TFIDF-RAG, Graph-RAG and Light-RAG.
\end{highlights}

\begin{keyword}
law article recommendation \sep Large Language Model \sep Knowledge Graph
\end{keyword}

\end{frontmatter}


\section{Introduction}
In a modern rule-of-law society, \textcolor{blue}{judicial  efficiency} is \textcolor{blue}{critical to safeguarding} economic growth and social stability\cite{dakolias2014court, Kap2024JudicialEfficiency, fix-fierro2003courts}. Meanwhile, \textcite{xvguang2024supreme} \textcolor{blue}{noted that courts currently face a massive case backlog}, \textcolor{blue}{resulting in a heavy workload for judicial personnel}. \textcite{teck4judge} \textcolor{blue}{emphasized an urgent need for intelligent tools} to assist judicial personnel in \textcolor{blue}{enhancing case adjudication efficiency}, \textcolor{blue}{as they boost judicial efficiency by categorizing similar cases, accelerating legal research, and keeping abreast of new laws—thus ensuring faster and more consistent decisions}.

Law article recommendation \textcolor{blue}{involves} predicting the relevant law articles applicable to a case based on its factual description. With the \textcolor{blue}{development} of computer science, \textcolor{blue}{numerous} data science techniques have been applied to this task, \textcolor{blue}{significantly enhancing} efficiency. Classic law article recommendation techniques—\textcolor{blue}{including} LSTM\cite{hochreiter1997long}, Text-CNN\cite{kim2014convolutional}, GRU\cite{cho2014learning}, RCNN\cite{lai2015recurrent}, HAN\cite{yang2016hierarchical}, and DPCNN\cite{johnson2017deep}—are based on traditional machine learning methods. These methods focus \textcolor{blue}{exclusively} on the correspondence between case facts and law article \textcolor{blue}{IDs}, while \textcolor{blue}{neglecting the semantic information within} law articles. \textcolor{blue}{Furthermore}, these methods are highly \textcolor{blue}{vulnerable to} insufficient data and imbalanced data labels. \textcolor{blue}{Thus, there is an urgent need to address these issues}.

\textcolor{blue}{Several} researchers \textcolor{blue}{have employed} Large Language Models (LLMs) \textcolor{blue}{for} law article recommendation\cite{yang2024large,ma2024leveraging,shui2023comprehensive}. \textcolor{blue}{With} large LLMs, users can not only retrieve law articles in \textcolor{blue}{numerical} and textual \textcolor{blue}{forms} but also \textcolor{blue}{explore the rationale for their applicability to cases through} follow-up questions. \textcolor{blue}{Furthermore}, users \textcolor{blue}{can probe the rationale for not selecting} other analogous law articles. However, \textcite{dahl2024large} found that \textcolor{blue}{direct use of} LLMs is prone to hallucinations, \textcolor{blue}{producing inaccurate} law article information. \textcolor{blue}{Thus, it is essential to incorporate} a fully accurate legal information and case knowledge base for auxiliary analysis.

\textcolor{blue}{Beyond improving the underlying knowledge base, a parallel line of work shows that prompt engineering can itself be a powerful lever for steering LLM behaviour and mitigating hallucinations. Recent surveys on hallucinations in LLMs highlight that a substantial portion of erroneous outputs is prompt-sensitive and can be reduced by disciplined prompting strategies, such as explicit uncertainty instructions, decomposition prompts, and verification-style prompts \cite{huang2023hallucination,dang2025survey}. Within this line of work, chain-of-thought prompting and its self-consistency variant encourage models to externalize intermediate reasoning and aggregate over multiple reasoning paths, which generally leads to more reliable answers on complex reasoning and knowledge-intensive tasks \cite{wei2022chain,wang2022self}. Building on the same intuition that exposing and checking intermediate reasoning improves faithfulness, verification-oriented prompting schemes such as Chain-of-Verification \cite{dhuliawala2024cove} ask the model to first draft an answer, then generate targeted follow-up questions, answer them independently, and finally revise the initial answer based on the verified evidence, thereby significantly decreasing hallucination rates in open-domain question answering. Complementary techniques like SelfCheckGPT \cite{manakul2023selfcheckgpt} further exploit prompt-based self-correction by sampling alternative completions and detecting contradictions as a black-box mechanism for filtering hallucinated content. Taken together, these studies suggest that prompt engineering provides an essential, model-agnostic axis for hallucination mitigation that naturally complements both architectural advances and RAG-style grounding. Motivated by this perspective, we design CLAKG-grounded prompts with explicit role specifications and strict output constraints, so that the LLM is systematically nudged to rely on verifiable legal evidence rather than fabricating non-existent law articles.}

To mitigate \textcolor{blue}{hallucination}, researchers \textcolor{blue}{have introduced} Retrieval-Augmented Generation (RAG) technology to enhance \textcolor{blue}{LLMs’ generation capabilities} \textcolor{blue}{by} retrieving relevant information from external knowledge bases\cite{lewis2020retrieval, cai2022recent}. Currently, RAG technology has been applied in \textcolor{blue}{domains} such as code generation\cite{gou2024rrgcode}, autonomous driving\cite{dai2024vistarag}, and enterprise management\cite{oleary2024rise}. However, research on \textcolor{blue}{RAG-based} law article recommendation remains relatively underexplored. RAG \textcolor{blue}{consists of} two \textcolor{blue}{components}: retrieval and generation. In \textcolor{blue}{academic discourse}, greater attention is often \textcolor{blue}{paid to} \textcolor{blue}{RAG’s} retrieval component, \textcolor{blue}{while} the generation process is \textcolor{blue}{primarily} handled by LLMs. Classic retrieval methods include \textcolor{blue}{TF-IDF}-based retrieval\cite{aizawa2003information}, BM25-based retrieval\cite{robertson2009probabilistic}, and \textcolor{blue}{frozen BERT-based} retrieval\cite{borgeaud2022improving}. However, \textcolor{blue}{these} methods \textcolor{blue}{primarily} focus on the \textcolor{blue}{lexical} level and rarely \textcolor{blue}{leverage macro-semantic information} for matching.

To \textcolor{blue}{address this issue}, \textcite{chaudhri2022knowledge} proposed that \textcolor{blue}{Knowledge Graphs (KGs)—structured databases that use directed graph structures to store data}—are highly suitable \textcolor{blue}{as knowledge bases} for RAG technology. \textcite{pan2024unifying} \textcolor{blue}{noted that integrating} KGs with LLMs \textcolor{blue}{holds significant potential for} future research. \textcite{edge2024from} \textcolor{blue}{employed} LLMs to automatically construct \textcolor{blue}{KGs}, which are then used to enhance \textcolor{blue}{LLMs’ generation capabilities}. \textcolor{blue}{This approach achieves strong performance in terms of} answer comprehensiveness and diversity. However, KGs generated \textcolor{blue}{via} this \textcolor{blue}{approach} often have complex structures and lack a fixed schema, making them difficult to modify—\textcolor{blue}{for example, adding new cases or removing outdated ones}.

\textcolor{blue}{To address the above issues, we propose to combine knowledge graph technology with large language models to improve the accuracy and interpretability of law article recommendation. We first pre-design the schema of a Case-Enhanced Law Article Knowledge Graph (CLAKG) that can simultaneously store current law articles and historical case information in a task-oriented manner. On this basis, we develop an automated CLAKG construction pipeline driven by LLMs. Furthermore, we build a closed-loop human–machine collaboration mechanism in which CLAKG supports retrieval-augmented prompting for LLMs, while expert feedback is continuously injected back into CLAKG to refine its structure.}

\textcolor{blue}{To assess the effectiveness of the proposed framework, we focus on Chinese criminal law as a representative civil-law domain and construct a dataset of criminal judgments from China Judgments Online. We compare our method with strong text-classification baselines based on BERT\cite{borgeaud2022improving}, GRU\cite{cho2014learning}, and DPCNN\cite{johnson2017deep}, as well as several RAG approaches such as TFIDF-RAG\cite{aizawa2003information}, Graph-RAG\cite{edge2024from}, and Light-RAG\cite{guo2024lightragsimplefastretrievalaugmented}. We further conduct ablation studies to investigate the contribution of different components of our framework.}

\textcolor{blue}{Building on the above design and empirical study, this work makes the following main contributions:}

\begin{enumerate}
  \item We propose a Case-Enhanced Law Article Knowledge Graph (CLAKG) that unifies law articles and adjudicated cases under a task-oriented schema, together with an automated CLAKG construction pipeline based on LLMs.

  \item We design an LLM--CLAKG-based law article recommendation framework that integrates CLAKG with LLMs in a retrieval-augmented generation (RAG) architecture. By grounding the LLM in CLAKG during recommendation, the framework mitigates hallucinations.

  \item We build a Chinese criminal law dataset and conduct extensive experiments showing that our method improves law article recommendation accuracy from 0.549 (LLM only) to 0.694 (LLM + CLAKG with case information), consistently outperforming strong text-classification baselines (BERT, GRU, DPCNN) and RAG baselines (TFIDF-RAG, Graph-RAG, Light-RAG).
\end{enumerate}

The remainder of this paper is organized as follows. Section~\ref{sec:relatedwork} reviews related work on law article recommendation, RAG technology, and knowledge graphs in the legal domain. Section~\ref{sec:Method} presents the proposed closed-loop law article recommendation framework. Section~\ref{sec:experiment} describes the construction of CLAKG and compares the accuracy of the proposed method with other baseline approaches. Lastly, Section~\ref{sec:conclusion} concludes the paper and outlines directions for future work.

\section{Related Work}
\label{sec:relatedwork}
\subsection{Law Article Recommendation} 
Early approaches to law article recommendation treated the task as a text classification or retrieval problem. Classical methods relied on keyword matching or manual feature engineering, but these struggled with the complex language of statutes. With the rise of deep learning, researchers applied neural models like LSTMs\cite{hochreiter1997long} and CNNs\cite{kim2014convolutional} to match case fact descriptions with relevant law articles. These models improved over bag-of-words baselines by capturing sequential and local context features. Transformer-based language models further advanced the field of BERT\cite{devlin2019bert} and its legal-domain adaptations (e.g., LegalBERT\cite{chalkidis2020legal}) enabled richer semantic representations of legal text. For example, Chalkidis et al.\cite{chalkidis2019neural} showed that a BERT-based classifier can predict applicable European human rights articles more accurately than traditional models, while Aletras et al.\cite{aletras2016predicting} demonstrated early on the feasibility of data-driven prediction of court decisions and their pertinent articles. Despite these gains, most prior models formulate law article recommendation as mapping facts to article IDs, without directly leveraging the content of the law articles themselves. This focus solely on fact-to-ID correspondence makes them heavily data-dependent and sensitive to label imbalance in training sets. In practice, it also limits interpretability – the models do not explain \emph{why} a recommended article is relevant, since the law articles are not explicitly used in the reasoning process. These limitations have motivated explorations into techniques that can incorporate external legal knowledge and provide better justification for recommendations.

\subsection{LLMs and Retrieval-Augmented Generation Techniques} 
The advent of LLMs, such as GPT-style transformers, has opened new avenues for law article recommendation. Rather than training a specific classifier, one can prompt an LLM with a case’s facts and ask for relevant law articles. Recent works have started to investigate this approach\cite{yang2024large,ma2024leveraging,shui2023comprehensive}, noting that LLMs can, in principle, leverage their vast pretrained knowledge to identify relevant law articles and even generate explanations. In particular, Shui et al.\cite{shui2023comprehensive} evaluate various state-of-the-art LLMs on legal judgment prediction subtasks (including relevant law article prediction), finding that while LLMs exhibit promising zero-shot capabilities, they still lag behind task-specific models on accuracy. A major concern with directly using LLMs for legal guidance is hallucination: the model may produce plausible-sounding but incorrect or nonexistent legal references. Dahl et al.\cite{dahl2024large} provide a sobering analysis of this issue, showing that even advanced LLMs frequently fabricate legal citations or statutes, which is unacceptable in high-stakes domains like law. To mitigate such hallucinations, researchers are turning to Retrieval-Augmented Generation (RAG) techniques. RAG combines LLMs with information retrieval: before or during text generation, the system retrieves relevant documents (e.g. law articles or case precedents) and conditions the LLM on that retrieved evidence. By grounding responses in actual law articles, RAG can significantly improve factual accuracy and trustworthiness of the model’s output. This approach was originally popularized in open-domain QA systems\cite{lewis2020retrieval} and has since been applied to legal NLP as well. For instance, a RAG-based legal assistant might fetch the text of candidate law articles given a case description, and the LLM then quotes or summarizes those law articles when recommending applicable law, thereby rooting its answer in verifiable sources. Key components of RAG include a retriever and a generator. In the legal domain, classic IR methods like TF–IDF\cite{aizawa2003information} or BM25\cite{robertson2009probabilistic} have been used to retrieve statute candidates by keyword overlap, while more recent approaches use neural retrievers (e.g. dense embeddings). Borgeaud et al.\cite{borgeaud2022improving} showed that even a frozen BERT-based retriever can substantially improve generation quality by supplying relevant text from a large corpus. However, naive retrieval based on surface text may miss conceptual matches. Legal phrasing varies, and a model might need to retrieve statutes using synonymous terms or implied connections, something neural retrieval aims to handle better than lexical methods. Overall, RAG methods represent a promising direction for law article recommendation, as they combine the reasoning ability of generative models with the grounding of information retrieval, thereby reducing hallucinations and improving the transparency of the recommendation.

\subsection{Knowledge Graphs in the Legal Domain} 
Beyond unstructured text retrieval, Knowledge Graphs (KGs) have emerged as valuable resources for legal information retrieval and recommendation. A legal knowledge graph represents entities such as law articles, cases, legal concepts, and their relationships (e.g., a case \emph{cites} a statute, or a chapter \emph{includes} certain articles) in a structured form. This graph structure can complement text-based methods by capturing the domain’s ontology and the interconnections between law articles. Prior works have constructed legal knowledge graphs to assist various tasks. For example, Chaudhri et al.\cite{chaudhri2022knowledge} argued that KGs are well-suited as the underlying database for RAG systems: instead of retrieving free text, one can query a knowledge graph to retrieve facts or linked nodes, ensuring that the generation is grounded in a consistent knowledge structure. In the legal domain, a KG could be queried to find statutes related to certain legal concepts or past cases with similar fact patterns. One advantage of KGs is that they store high-level associations beyond lexical similarity – for instance, a graph can link a section of law to broader categories or to precedent cases, enabling the recommendation system to recall relevant law articles even if a case’s description uses different wording. There is growing interest in integrating KGs with LLMs\cite{pan2024unifying}. Pan et al. \cite{pan2024unifying} outline a roadmap for unifying the strengths of symbolic knowledge graphs and distributed language models, noting that KGs can provide factual grounding for LLMs while LLMs can help populate and expand KGs from raw texts. In the context of law, researchers have begun designing hybrid systems that use KGs alongside LLMs. Edge et al.\cite{edge2024from} explore automatic construction of knowledge graphs using LLMs (generating graph triples from unstructured legal text) and using those graphs to enhance answer generation. This suggests a future where legal KGs and LLMs continuously enrich each other: the KG offers a source of truth to reduce LLM errors, and the LLM helps keep the KG up-to-date with new insights. The integration of knowledge graphs in legal AI is still an emerging area, and challenges remain, such as ensuring the completeness and currency of the graph. Nonetheless, the combination of KGs with retrieval-augmented LLMs holds great promise for building more reliable and interpretable law article recommendation systems in the coming years.\\

\section{Method}
\label{sec:Method}

\begin{figure*}[h]
\includegraphics [%
    width=1\textwidth] {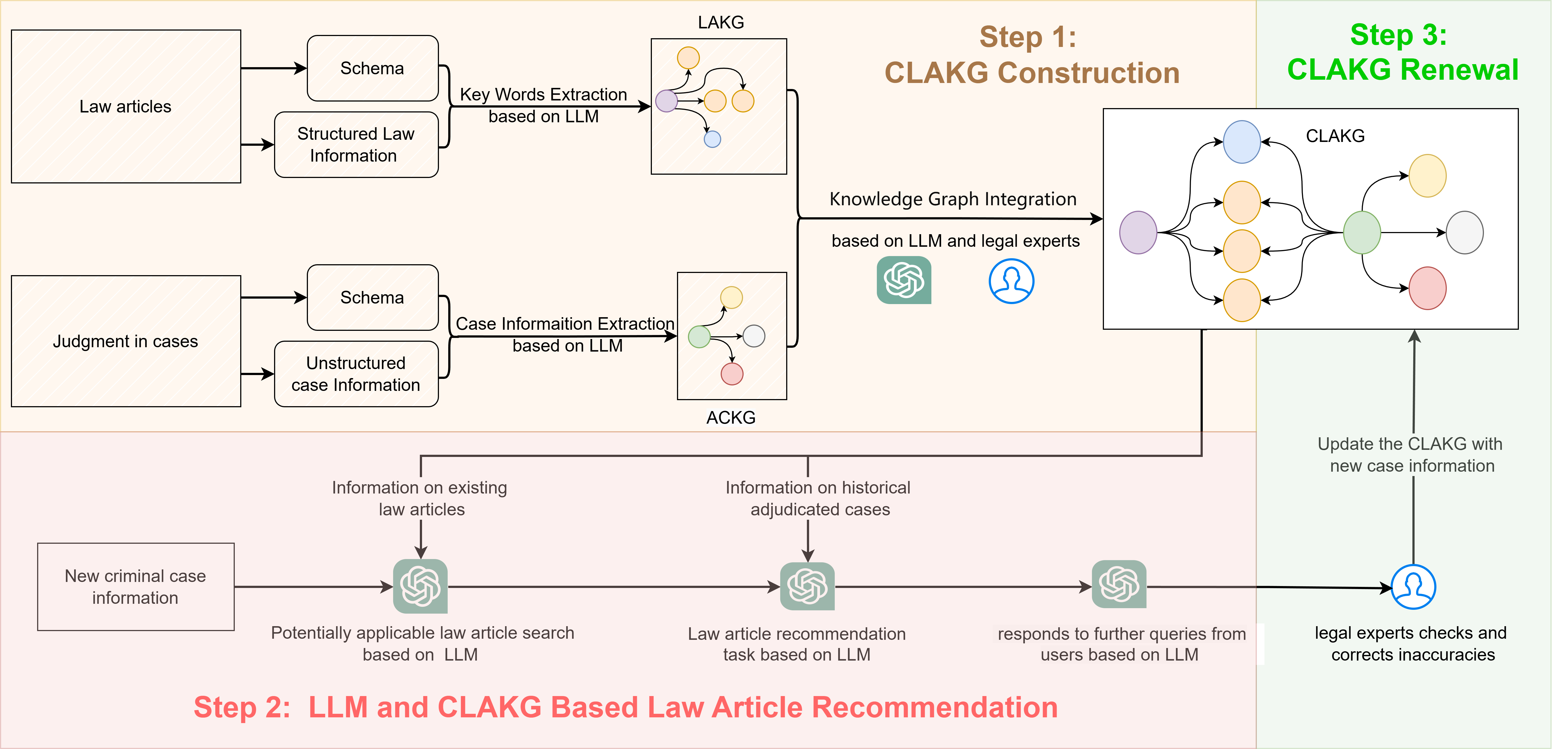}
\caption{\textcolor{blue}{Overall closed-loop pipeline of the proposed CLAKG-based law article recommendation system. The Case-Enhanced Law Article Knowledge Graph (CLAKG) is composed of the Law Article Knowledge Graph (LAKG) and the Adjudicated Cases Knowledge Graph (ACKG).}}
\label{fig:quanwenliuchengtu}
\vspace{-3mm}
\end{figure*}
\textcolor{blue}{The overall closed-loop pipeline of the proposed CLAKG-based law article recommendation system is shown in Figure \ref{fig:quanwenliuchengtu}. It consists of three main stages: (i) constructing the Case-Enhanced Law Article Knowledge Graph (CLAKG) (Section~3.1); (ii) performing LLM--CLAKG based law article recommendation (Section~3.2), as shown in Figure \ref{fig:LawRec}; and (iii) human--machine collaboration, where legal experts review and correct the recommended articles and the CLAKG is updated accordingly (Section~3.3). }

We first establish a knowledge graph based on the designed schema. Initially, nodes and relationships are extracted from law articles and judgments in cases to construct the Law Article Knowledge Graph (LAKG) and the Adjudicated Cases Knowledge Graph (ACKG). Subsequently, by utilizing a Large Language Model (LLM), these two graphs are integrated to form CLAKG. This part is introduced in Section \ref{sub:CPLKG_construction}. Building upon this, we have developed a human--machine collaborative law article recommendation framework. Users describe new case information, and the LLM provides potentially applicable law articles based on keyword matching and graph embedding techniques. Moreover, the system can provide similar historical case information from the CLAKG, followed by providing the most relevant law articles that match the new case using LLM (Section \ref{sub:law_article_rec_fw}). \textcolor{blue}{Following the above process, the legal expert reviews the law articles provided by the system and corrects any inaccuracies. The new case information is used to update the CLAKG, forming a closed-loop human–machine collaboration process whose practical effect is illustrated in Section \ref{sec33}.}

\subsection{Case-Enhanced Law Article Knowledge Graph Construction}
\label{sub:CPLKG_construction}

\begin{figure}[h]
\centering
    \includegraphics[width=1\textwidth]{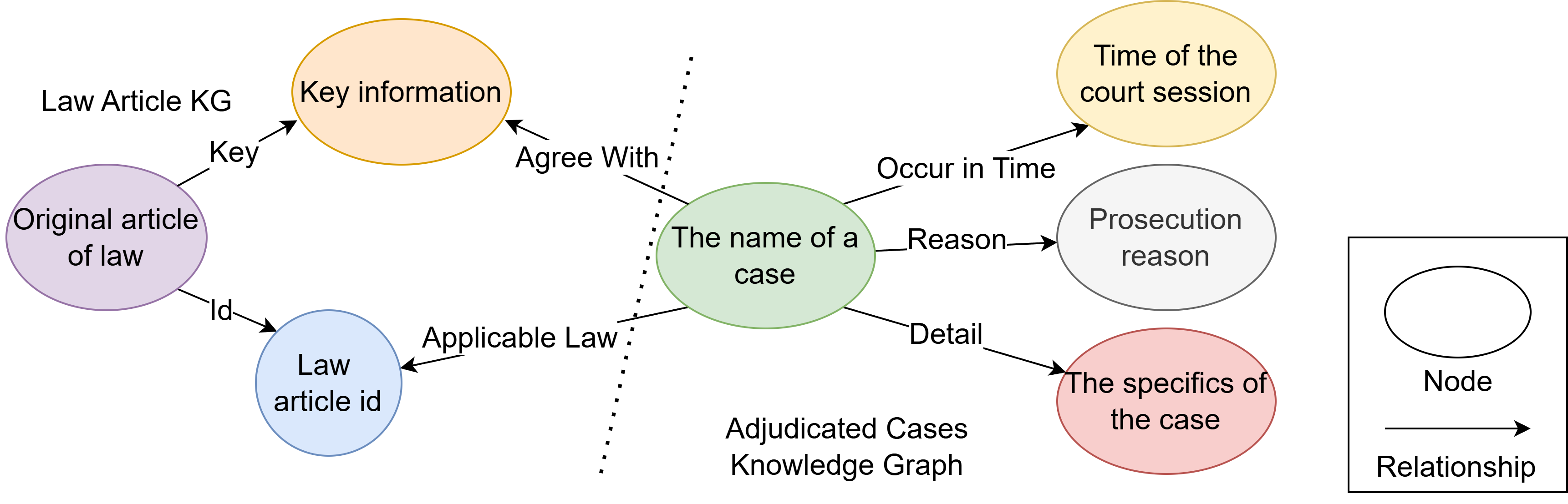}
    \centering
    \caption{\textcolor{blue}{Schema of CLAKG. Ellipses of different colors represent different entity types.}}
    \label{fig:clakg_schema}
\end{figure}

\begin{figure}[h]
\centering
\includegraphics[width=1\textwidth]{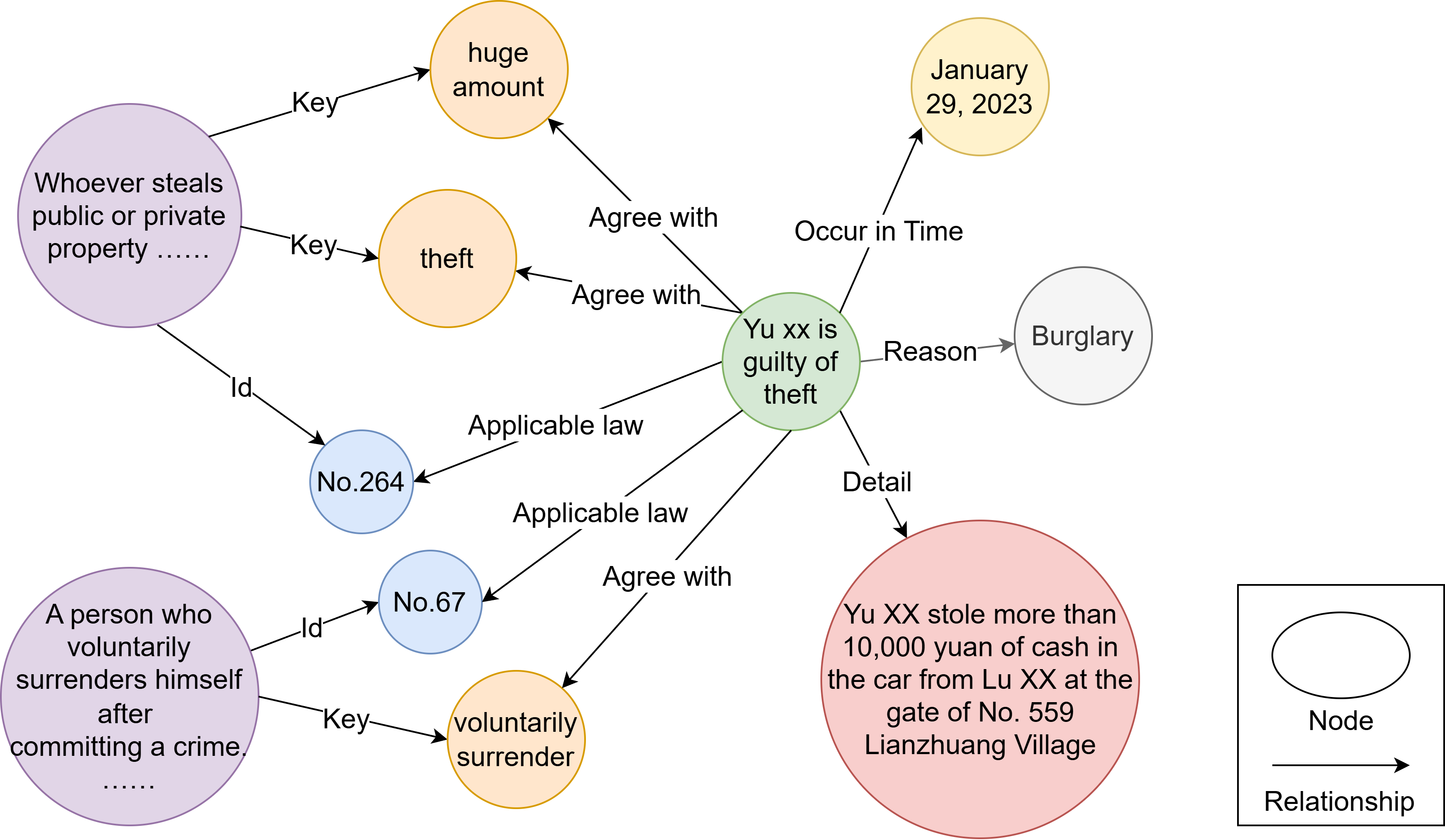}
\caption{\textcolor{blue}{Example subgraph of CLAKG instantiated according to the schema defined in Figure~2.}}
\centering
\label{fig:clakg_example}
\vspace{-3mm}
\end{figure}
\begin{table}[h]
\caption{Entity types in Case-Enhanced Law Article Knowledge Graph}
\vspace{-5mm}
\label{table5}
\begin{center}
\begin{tabular}{m{0.3\textwidth}|m{0.65\textwidth}}
\textbf{Entity Type}  & \textbf{Description}  \\
\hline Original article of law &  \setlength{\baselineskip}{10pt}
An original article from the field of law, such as criminal law or traffic law, etc
\\ \hline
 Key information & 
Key information in law article 
 \\ \hline
 Law article id & The article number in the law \\ \hline
The name of a case &  
Case name on the judgment documents 
 \\ \hline
 Time of the court session & 
The time when a court makes a decision in a case
\\ \hline
 Prosecution reason & 
Prosecution reason on the judgment documents
\\ \hline
 The specifics of the case & 
Case information extracted using a large 
language model
\\ 
\end{tabular}
\end{center}
\vspace{-2mm}
\end{table}

\begin{table}[h]
\caption{Relationship types in Case-Enhanced Law Article Knowledge Graph}
\label{table6}
\vspace{-5mm}
\begin{center}
\begin{tabular}{p{0.25\textwidth}|p{0.7\textwidth}}
\textbf{Relation Type}  &  \textbf{Description}  \\
\hline Key & 
A law article contains several pieces of key information
 \\ \hline
 Id & 
A law article has a unique law article id corresponding to it
 \\ \hline
 Agree With & 
A case agrees with several pieces of key information
 \\ \hline
Applicable Law & The law that matches the case
 \\ \hline
 Occur in Time & 
The verdict in a case occurs on a fixed date
 \\ \hline
 Reason & 
Links a case to its prosecution reason stated in the judgment
 \\ \hline
 Detail & 
Links a case to its fact description
 \\
\end{tabular}
\end{center}
\vspace{-8mm}
\end{table}
\textcolor{blue}{Figure \ref{fig:clakg_schema} illustrates the schema of CLAKG, defining the entity types and relation types used in our system. An example subgraph instantiated according to this schema is shown in Figure \ref{fig:clakg_example}.  Tables \ref{table5} and \ref{table6} provide the details of all types of nodes and relationships in the CLAKG.}

As shown in Figure~\ref{fig:clakg_schema}, the Case-Enhanced Law Article Knowledge Graph (CLAKG) is composed of the Law Article Knowledge Graph (LAKG) and the Adjudicated Cases Knowledge Graph (ACKG). The LAKG includes original article of law, law article id, and key information. The key information is first extracted from the original article of law using an LLM, and then reviewed and, if necessary, refined by legal experts before being stored as nodes in CLAKG. The ACKG contains the name of a case, time of the court session, prosecution reason, and the specifics of a case. The description of the case process is derived from the content of the judgment documents processed using an LLM, and again experts check and adjust the extracted elements so that the resulting case nodes and their attributes faithfully reflect the original judgments.\\

\textcolor{blue}{Based on the reference law articles cited in the judgment documents, we link each name of a case node in ACKG to the corresponding law article id nodes. Furthermore, we analyze the correlation between the factual description of a case and the key information of the applicable law articles using an LLM, and select up to five of the most relevant key-information nodes for each case. Concretely, the LLM is asked to choose several items from the existing list of key-information strings for the cited articles; its answer is parsed as a semicolon-separated list and each substring is matched against this list. Any substring that does not exactly match an existing key-information node is discarded, and again no new nodes are added. After this automatic step, legal experts can confirm the suggested case–article links and key-information associations, correct inaccurate links, and add missing but legally important relations if the LLM has overlooked them. This human–machine collaborative filtering strategy ensures that the constructed CLAKG remains consistent with the predefined schema: noisy LLM outputs may reduce recall before expert correction, but they cannot introduce spurious concepts or inconsistent node types into the graph. Finally, each name of a case node is linked to several key information nodes.}


\subsection{LLM and CLAKG Based Law Article Recommendation Framework}
\label{sub:law_article_rec_fw}
\begin{figure}[h]
\centering
\includegraphics [%
    width=1\textwidth] {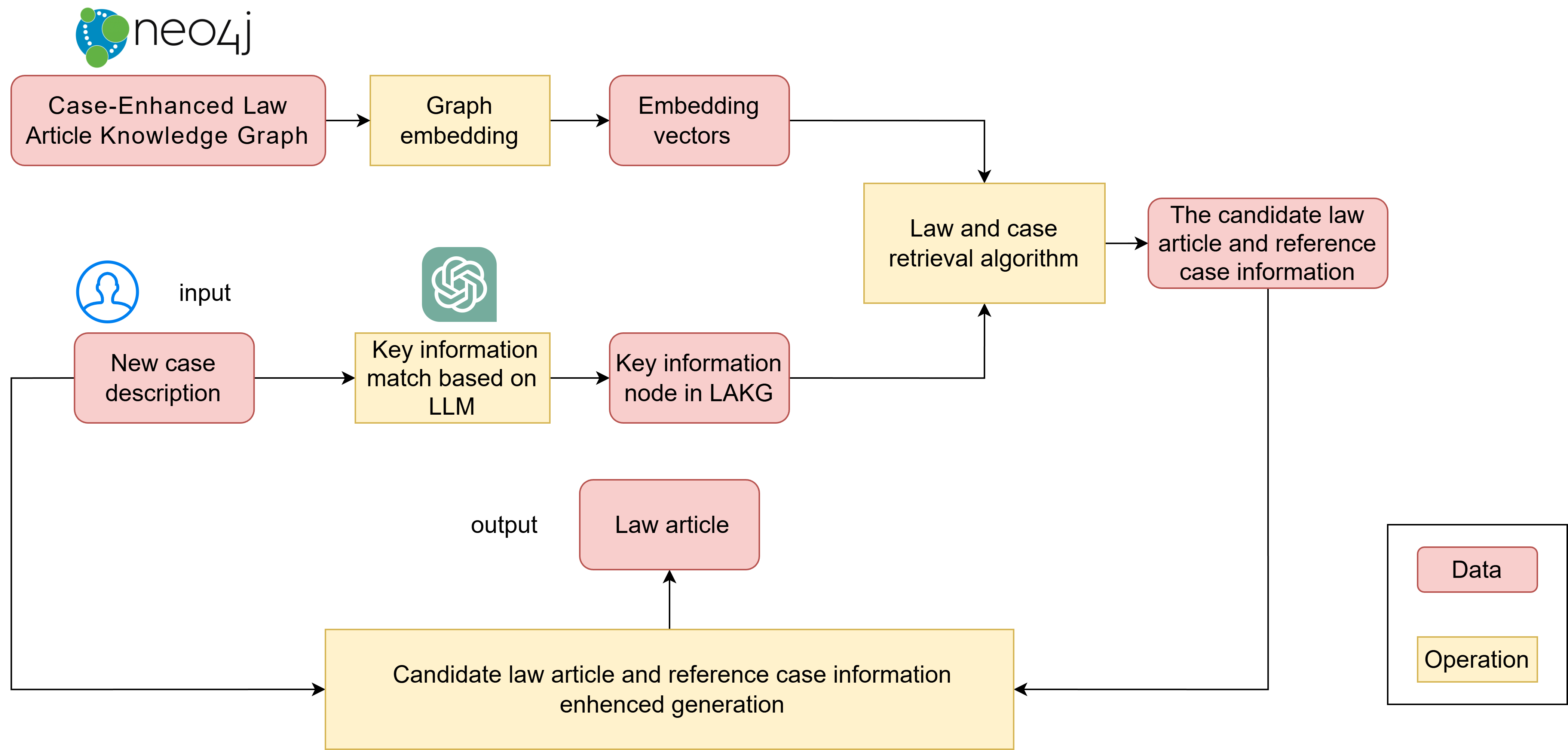}
\caption{\textcolor{blue}{The LLM and CLAKG based law article recommendation framework. The candidate law articles and historical cases retrieved from CLAKG are used to constrain the LLM’s reasoning and thereby mitigate hallucinations.}}
\label{fig:LawRec}
\vspace{-3mm}
\end{figure}
The LLM and CLAKG based law article recommendation framework is shown in Figure \ref{fig:LawRec} \textcolor{blue}{and step 2 of Figure \ref{fig:quanwenliuchengtu}.} Before the start of the law article recommendation task, the CLAKG constructed in Section \ref{sub:CPLKG_construction} needs to undergo graph embedding preprocessing using the Relational Graph Convolutional Network (RGCN) model mentioned in Section 3.2.1. The process of law article recommendation begins with the user inputting a newly emerged case. In Section 3.2.2, a LLM is used to match the K most relevant keywords to this new case (with K=8 as selected in this paper). Based on this, in Section 3.2.3, a candidate law article retrieval algorithm is employed to retrieve q candidate law articles that may be suitable for the new case (with q=5 as selected in this paper). In Section 3.2.4, the information from the new case input by the user is integrated with the candidate law articles and historical case information obtained in Section 3.2.3 into a prompt, which is then passed to the LLM to complete the law article recommendation task, outputting at most $t$ law article IDs (with $t=1$ as selected in this paper).

\subsubsection{Graph Embedding on CLAKG with RGCN}
\label{rgcn}

\textcolor{blue}{Given the CLAKG constructed in Section~\ref{sub:CPLKG_construction}, we now describe how to compute node embeddings on this graph. Let $G = (V, E, R)$ denote the CLAKG, where each node $v \in V$ belongs to one of the entity types listed in Table~1 (original article of law, law article id, key information, case name, time of the court session, prosecution reason, and case details). Edges $e \in E$ are labeled by relation types $r \in R$ from Table~2, including \emph{Key} (linking an article to its key information), \emph{Id} (linking an article to its law article id), \emph{Applicable Law} (linking a case to the articles cited in the judgment), \emph{Agree With} (linking a case to the key information it matches), \emph{Occur in Time} (linking a case to its decision date), \emph{Reason} (linking a case to the reason of lawsuit), and \emph{Detail} (linking a case to its detailed fact description). Together, these nodes and relations encode which law articles apply to which cases and through which key information, and they constitute the structural basis on which our graph embedding module operates.}

\textcolor{blue}{The purpose of the graph embedding module in our framework is to exploit the heterogeneous, multi-relational structure of CLAKG and map every node $v \in V$ to a low-dimensional vector representation. These embeddings provide the matching capability in our framework: in Section~\ref{sub:law_article_rec_fw}, they are used to measure the similarity between the key-information nodes selected from a new case and all law-article nodes when ranking candidate law articles. To learn such representations on CLAKG, we adopt Relational Graph Convolutional Networks (RGCN)~\cite{schlichtkrull2018modeling}, which extend standard Graph Convolutional Networks (GCNs)~\cite{kipf2016semi} to multi-relational graphs by assigning different transformation matrices to different relation types and aggregating information from neighbors according to the graph structure.}

The node embedding vector update formula for the RGCN model is shown in Equation \ref{equa_1}.\\
\begin{equation}
\label{equa_1}
h_i^{(l+1)}=\sigma\left(\sum_{r \in \mathcal{R}} \sum_{j \in \mathcal{N}_i^r} \frac{1}{c_{i, r}} W_r^{(l)} h_j^{(l)}+W_0^{(l)} h_i^{(l)}\right)
\end{equation}
where $h_i^{(l)}$ represents the embedding vector of node $i$ in the $l$-th layer, $N_i^r$ represents all neighboring nodes of node $i$ that have relationship $r$. $c_{i, r}$ is a normalization constant, which in this paper is taken as $\left|N_i^r\right|$. Both $W_r^{(l)}$ and $W_0^{(l)}$ are parameter matrices.

The graph embedding training task adopted in this paper is link prediction. For the link prediction task, part of the dataset consists of several triples (head node, relation, tail node) selected from the knowledge graph $G=(\mathcal{V}, \mathcal{E}, \mathcal{R})$, which are used as positive examples. Another part of the dataset consists of randomly selected entities and relations from the knowledge graph, which are used to construct triples (head node, relation, tail node) that do not actually have a connection in the graph, i.e., fabricated triples, used as negative examples. The training objective of the model is to distinguish whether the given triples truly exist in the graph or are fabricated. The number of positive and negative triples is equal, which helps to prevent bias during model training.\\
\begin{figure}[h]
\centering
\includegraphics [%
    width=1\textwidth] {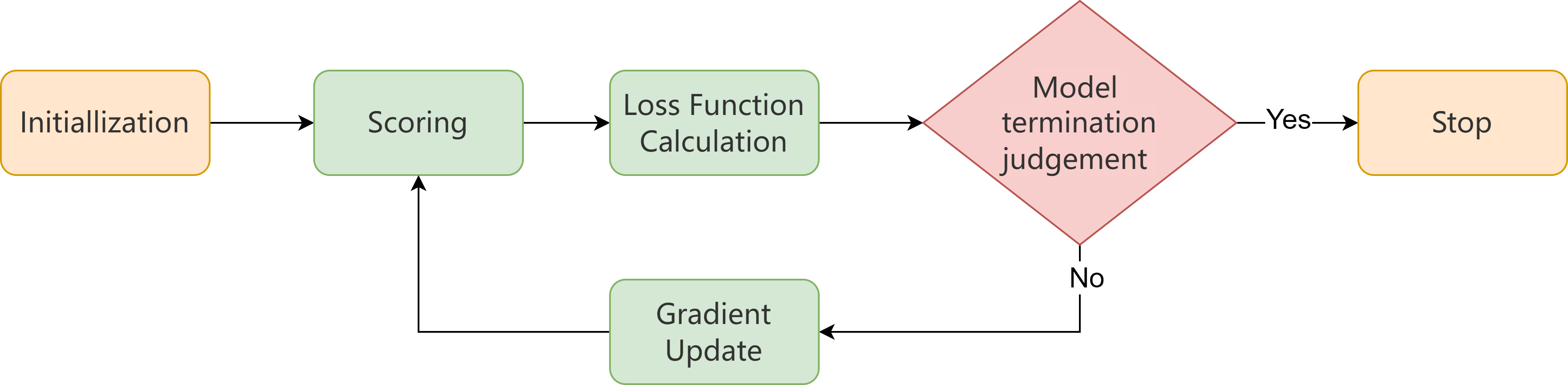}
\caption{
Flowchart of RGCN training process.}
\label{fig:RGCNmodel}
\end{figure}

The training process is illustrated in Fig.~\ref{fig:RGCNmodel}. We first randomly initialize trainable embedding vectors for all entities and relations in the knowledge graph. Node representations are then updated by stacking $L$ layers of relational graph convolution (RGCN) as in Eq.~\ref{equa_1}, where each layer aggregates messages from relation-specific neighbors with appropriate normalization (e.g., by the neighborhood size) and applies a non-linear activation.

Given the final node embeddings, we compute the plausibility score of each triple $(s,r,o)$ using the DistMult \cite{yang2014embedding} factorization:
\begin{equation}
\label{equa_2}
f(s,r,o)=\langle e_s, e_r, e_o\rangle
=\sum_{k=1}^{d} e_{s,k}\,e_{r,k}\,e_{o,k},
\end{equation}
where $e_s,e_o\in\mathbb{R}^{d}$ denote the embeddings of the subject node $s$ and object node $o$, respectively, and $e_r\in\mathbb{R}^{d}$ is the embedding of relation $r$.

We optimize the model using the binary cross-entropy (BCE) loss in Eq.~\ref{eq:loss_bce_unweighted}. Let $\mathcal{T}$ denote the set of training examples consisting of observed positive triples and negatively sampled (corrupted) triples generated by replacing the head or tail entity with a randomly selected entity (i.e., head/tail corruption). The binary label $y\in\{0,1\}$ indicates whether a triple is positive ($y=1$) or a negative sample ($y=0$):
\begin{equation}
\label{eq:loss_bce_unweighted}
\mathcal{L}
= -\frac{1}{|\mathcal{T}|}
\sum_{(s,r,o,y)\in\mathcal{T}}
\Big[
y \log \sigma(f(s,r,o))
+ (1-y)\log\!\big(1-\sigma(f(s,r,o))\big)
\Big],
\end{equation}
where $\sigma(\cdot)$ is the logistic function defined in Eq.~\ref{equa_4}.

\begin{equation}
\label{equa_4}
\sigma(x)=\frac{1}{1+e^{-x}}=\frac{e^x}{e^x+1}.
\end{equation}

\subsubsection{Matching Key Information Nodes in CLAKG with LLM}
\label{subsec322}
\begin{table}[h]
\caption{\textbf{Prompts for LLM in the task of matching key information nodes in the CLAKG related to new case information.}}
\label{match_key_inf}
\setlength{\tabcolsep}{4pt}
\begin{tabular}{m{0.3\textwidth}|m{0.7\textwidth}}
 \textbf{Component} & \textbf{Details} \\
\hline Expert Role & Expert in law article analysis \\
\hline Task Description & 
Given the available new case information and the 
following key information nodes, please output $0-8$ key
information nodes most relevant to the case. If you 
believe there are fewer than 8 relevant key information 
nodes, there is no need to force the list to reach 8 nodes.
 \\
\hline New Case Information & Description of the new case input by the user \\
\hline Key Information Nodes & all key information nodes in the CLAKG\\ 
\hline Precautions & 
Please directly output the key information nodes, separated 
by semicolons (;) without including any additional content 
(including unnecessary punctuation marks).
 \\
\hline Output Example & 
Public and private property; sale; committing 
a crime; causing damage; seriously disrupting public order; 
multiple thefts; lawfully performing duties
 \\
\end{tabular}
\end{table}
In this section, we aim to match the relevant key information nodes within CLAKG based on current case information provided by users. To facilitate this process, we have designed prompts for the LLM, which include the expert role, task description, details of the newly reported case, known key information nodes, and example outputs. The prompt template for LLM in the task of matching key information nodes in the CLAKG related to new case information is shown in Table \ref{match_key_inf}.

\textcolor{blue}{During inference, we reuse the global list of key-information strings stored in CLAKG and present it to the LLM as the option set in the prompt. The LLM is instructed to output several items from this list, separated by semicolons. The output is post-processed by splitting on semicolons and performing exact matching against the list; substrings that do not match any existing key-information node are discarded and do not create new nodes.}

\subsubsection{Law Article and Case Retrieval Algorithm}
\label{subsec323}
\begin{algorithm*}[h!]
\caption{Candidate Law Article Retrieval Algorithm}
\label{algorithm1}
\begin{algorithmic}[1]
\Require Query key-information nodes $K$; embedding lookup table $\mathrm{emb}[\cdot]$ (Section~3.2.1); relation type $\{\textit{Key}\}$ in CLAKG
\Ensure Top-q candidate $\{\textit{Original article of law}\}$ nodes (optionally with scores)

\State \textbf{Initialize:} $S \gets \emptyset$; $\mathcal{L} \gets [\ ]$ 
\Comment{$S$: candidate law-article set, $\mathcal{L}$: list of (score, law\_article)}

\For{$e \in K$}
    \State $S \gets S \cup \{\, x \mid (x \xrightarrow{\{\textit{Key}\}} e) \in E \ \wedge\ x \in \{\textit{Original article of law}\} \,\}$
\EndFor

\For{$a \in S$}
    \State $s \gets 0$
    \For{$e \in K$}
        \State $s \gets s + \cos(\mathrm{emb}[a], \mathrm{emb}[e])$
        \Comment{$\cos(u,v) = \frac{u \cdot v}{\|u\|\|v\|}$}
    \EndFor
    \State $\mathcal{L}.\texttt{append}((s, a))$
\EndFor

\State \texttt{sort}($\mathcal{L}$) by $s$ in descending order
\State \Return first q law articles in $\mathcal{L}$ (or their IDs)
\end{algorithmic}
\end{algorithm*}

In this section, we propose an algorithm for retrieving historical case information and candidate law articles by leveraging the aforementioned data to identify the law article most relevant to the current case. 

The proposed method for retrieving candidate law article involves traversing all `Key information’ nodes obtained in Section 3.2.2 and retrieving the `Original Article of Law' nodes that are connected to the `Key Information’ nodes via `Key’ relationships. These nodes are then aggregated into a set container, with all `Original article of law' nodes in the set representing the candidate law article. Given that the number of candidate law articles in the set container may be substantial, it is essential to select those law articles most analogous to the newly reported case for further analysis. To this end, each node in the container is evaluated by calculating the similarity scores (cosine similarity) between the `Original article of law nodes’ embedding vector and the multiple `key information’ nodes embedding vector obtained in 3.2.1. 

These similarity scores are summed to yield a cumulative similarity score between each `Original Article of Law' node and the newly reported case. Ultimately, the top five `Original Article of Law' nodes with the highest cumulative similarity scores are selected as the candidate law articles, serving as reference points for the LLM in completing the law article recommendation task.

The pseudo-code is shown in Algorithm \ref{algorithm1}. The time complexity of this algorithm is \( O(m \cdot n) + O(m \log m) \), where \( m \) is the number of elements in the \textit{law article set} and \( n \) is the number of elements in the \textit{key information node} set.

After the above process, by querying the CLAKG for nodes representing `the name of a case' connected to these law articles, we can obtain historical case information for reference.

\subsubsection{LLM-Enhanced Law Article Recommendation}
\label{subsec324}

\begin{figure}[h]
\centering
\includegraphics [%
    width=1\textwidth] {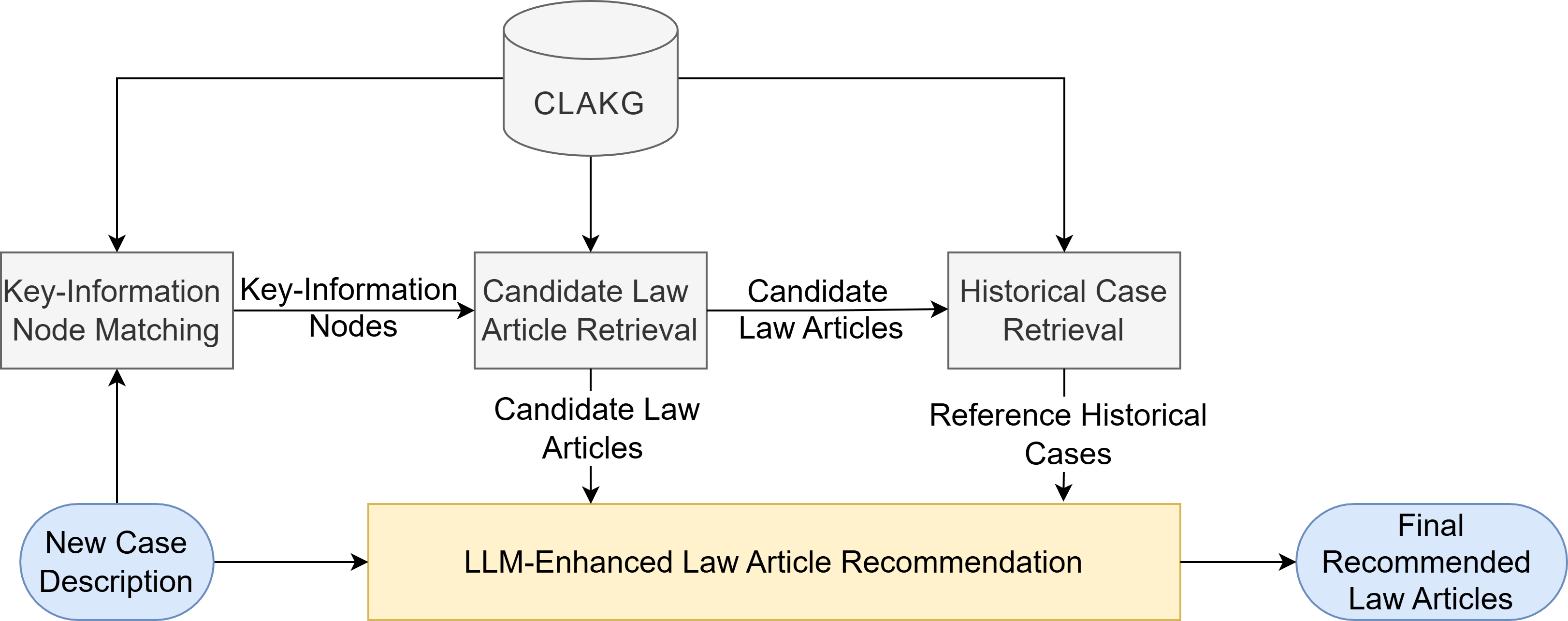}
\caption{\textcolor{blue}{Flowchart of LLM-enhanced law article recommendation grounded in CLAKG. The LLM receives KG-retrieved candidate law articles and linked historical cases as verifiable context, which constrains its reasoning and helps mitigate LLM hallucinations.}}
\label{fig:llm_integration}
\vspace{-3mm}
\end{figure}

\textcolor{blue}{As illustrated in Figure~\ref{fig:llm_integration}, once the CLAKG-based retrieval stages in Sections~3.2.2 and 3.2.3 have been completed, the system has obtained three types of structured evidence for a new case: (i) the textual description of the new case provided by the user, (ii) a small set of candidate law articles retrieved from CLAKG, and (iii) a set of reference historical cases linked to these candidate law articles. In this subsection we describe how this information is integrated into a Large Language Model so as to perform law article recommendation while mitigating hallucinations.}

\textcolor{blue}{To explicitly leverage the retrieved candidate law articles and reference historical cases as constraints for mitigating hallucinations, we design a structured prompt for the LLM-enhanced recommendation stage. For each new case, the prompt specifies the expert role, task description, the factual description of the case, the retrieved candidate law articles and their matched key-information nodes, as well as the reference historical cases linked to these articles. In addition, we impose strict output constraints that restrict the model to selecting from the candidate set and require it to explicitly state that no suitable law article is found rather than fabricating new ones. The detailed prompt components for CLAKG-grounded law article recommendation are summarized in Table~\ref{tab:llm_rec_prompt}.}

\begin{table}[h]
\caption{\textcolor{blue}{Prompts for LLM in the task of CLAKG-grounded law article recommendation.}}
\label{tab:llm_rec_prompt}
\setlength{\tabcolsep}{4pt}
\begin{tabular}{m{0.3\textwidth}|m{0.7\textwidth}}
\textbf{Component} & \textbf{Details} \\
\hline
Expert Role & Criminal-law expert specializing in applying Chinese criminal law articles to concrete cases. \\
\hline
Task Description & 
Given the new case description, the candidate law articles, and the reference historical cases, recommend the law articles that are most applicable to the case. You must choose \emph{only} from the candidate law articles and must not fabricate non-existent article numbers or law articles. \\
\hline
New Case Description & 
Factual description of the new case provided by the user. \\
\hline
Candidate Law Articles & 
For each candidate article: its ID and original article of law. \\
\hline
Reference Historical Cases & 
Concise summaries of historical cases linked to each candidate article in CLAKG (the specifics of the case, prosecution reason, and judgment result). \\
\hline
Output Constraints & 
Output at most $t$ law article IDs from the candidate set, each with a brief justification. If none of the candidates is suitable, explicitly state that no applicable law article is found, rather than inventing a new one. \\
\end{tabular}
\end{table}

\textcolor{blue}{Given this prompt, the LLM compares the new case with each candidate law article and its linked historical cases, and then produces a ranked list of recommended law articles together with short textual explanations. The law articles remaining after these checks are returned to the user as the \emph{Final Recommended Law Articles} in Figure~\ref{fig:llm_integration}. The user can review these recommendations, request further explanations from the LLM (e.g., why a particular article applies or why similar articles are not chosen).}

\subsection{\textcolor{blue}{Closed-loop Human--Machine Collaboration and CLAKG Renewal}}
\label{sec33}

\begin{algorithm}[h]
\small
\caption{\textcolor{blue}{CLAKG Update with Human Feedback}}
\label{algorithm_update}
\begin{algorithmic}[1]
\Require CLAKG $G=(V,E,R)$, new case $c$, system law-article recommendations $A^{\text{sys}}$, expert-validated law articles $A^{\text{exp}}$, flag \textit{accepted}
\Ensure Updated CLAKG $G'=(V',E',R)$
\State $V' \gets V,\; E' \gets E$
\If{$c \notin V'$} add $c$ to $V'$ \EndIf
\State add / update case-detail nodes for $c$ and link them to $c$
\State obtain $K^{\text{sys}}(c)$ via the LLM-based keyword matching in Section~3.2.2 and expert-validated $K^{\text{exp}}(c)$

\If{\textit{accepted} = \textbf{True}}  \Comment{recommendation accepted}
    \State $E' \gets E' \cup \{(c,a,\texttt{Applicable Law}) \mid a \in A^{\text{exp}}\} \cup \{(c,k,\texttt{Agree With}) \mid k \in K^{\text{exp}}(c)\}$
\Else                                        \Comment{recommendation corrected}
    \State $E' \gets E' 
      \setminus \{(c,a,\texttt{Applicable Law}) \mid a \in A^{\text{sys}}\} 
      \setminus \{(c,k,\texttt{Agree With}) \mid k \in K^{\text{sys}}(c)\}$
    \State $E' \gets E' 
      \cup \{(c,a,\texttt{Applicable Law}) \mid a \in A^{\text{exp}}\} 
      \cup \{(c,k,\texttt{Agree With}) \mid k \in K^{\text{exp}}(c)\}$
\EndIf
\State \Return $G'=(V',E',R)$
\end{algorithmic}
\end{algorithm}

\textcolor{blue}{In practical deployment, the LLM--CLAKG framework operates in a closed-loop manner in collaboration with legal experts, as shown in step 3 of Figure~\ref{fig:quanwenliuchengtu} and formalized in Algorithm~\ref{algorithm_update}. After the system has produced a recommended law article and a natural-language explanation for a newly emerged case (Section~\ref{sub:law_article_rec_fw}), experts review the proposed result. This review focuses on both the correctness of the recommended article and the adequacy of the supporting reasoning. In parallel, the system also outputs a small set of key-information nodes (``keywords'') that it deems most relevant to the new case, obtained by reusing the LLM-based matching procedure in Section~3.2.2. Experts inspect and, if necessary, adjust this keyword set as well, so that both the applicable law articles and their supporting key information are explicitly validated.}

\textcolor{blue}{If the recommendation is confirmed to be appropriate, the corresponding case--law article pair is treated as a new, high-quality supervision signal for the knowledge graph. Concretely, the case node of the current judgment is added to CLAKG, and one or more ``Applicable Law'' edges are created between this case node and the validated law-article nodes. In addition, the occurrence time, prosecution reason, and a concise description of the case facts stored as case-detail nodes, are linked to the new case node via ``Occur in Time'', ``Reason'' and ``Detail'' relationship. Building on the expert-validated keyword set, the system then connects the new case to the corresponding key-information nodes by adding ``Agree With'' edges, again strictly reusing only key-information nodes that already exist in CLAKG. These operations enrich CLAKG with up-to-date case--article and case--keyword links without modifying the underlying LLM or the retrieval algorithm, corresponding to the ``accepted'' branch in Algorithm~\ref{algorithm_update}.}

\textcolor{blue}{Conversely, if experts deem the recommendation partially or completely incorrect, they can either select the correct articles from the candidate list or manually specify the appropriate law articles. The system then updates CLAKG accordingly: erroneous ``Applicable Law'' edges can be removed, while new edges are established between the case node and the expert-validated law articles. At the same time, any misleading keyword associations proposed by the system are pruned by removing the corresponding ``Agree With'' edges, and new ``Agree With'' edges are created between the case node and the expert-corrected key-information nodes. This correction process is captured by the ``corrected'' branch in Algorithm~\ref{algorithm_update}, where expert feedback directly refines both the case--law article and case--keyword structure of the graph.}

\textcolor{blue}{This human--machine collaboration enables CLAKG to continuously absorb new cases, updated keyword associations, and law article changes, leading to a denser and more accurate graph. Since subsequent recommendations always start from CLAKG-based retrieval and KG-grounded prompting, an improved graph directly translates into better keyword matching in Section~3.2.2, more precise candidate law-article retrieval in Section~3.2.3, and stronger factual grounding for the LLM in Section~3.2.4. As a result, the closed-loop renewal of CLAKG---covering both law-article and key-information links for new cases---gradually improves recommendation quality and helps mitigate hallucinations in future LLM outputs.}

\section{Experiment}
\label{sec:experiment}

\subsection{Datasets}

\textcolor{blue}{While the general architecture of our framework is designed to be not limited to a specific jurisdiction in principle, in this study we instantiate and evaluate it on Chinese criminal law as a concrete case study.} Specifically, the law-article dataset employed in this section is derived from the Amendment (XI) to the Criminal Law of the People’s Republic of China \cite{NPC2020}, ratified during the 24th session of the Standing Committee of the 13th National People’s Congress. This amendment is systematically structured into two sections: the General Provisions and the Specific Provisions, encompassing a total of 452 articles—101 in the General law articles and 351 in the Specific law articles. In other words, our law article set already covers the \emph{entire} universe of criminal law articles under Amendment (XI), and the number of law-article nodes in CLAKG directly reflects the full statutory space relevant to this domain; it cannot be meaningfully “expanded” without moving beyond Chinese criminal law itself. The CLAKG was constructed utilizing the methodology in Section~\ref{sub:CPLKG_construction}. \textcolor{blue}{We discuss in Section~\ref{sec:conclusion} the challenges and potential of adapting this framework to other legal systems (e.g., common-law jurisdictions) and languages.}\\

\begin{figure}[h]
\vspace{-5mm}
\includegraphics [%
    width=1\textwidth] {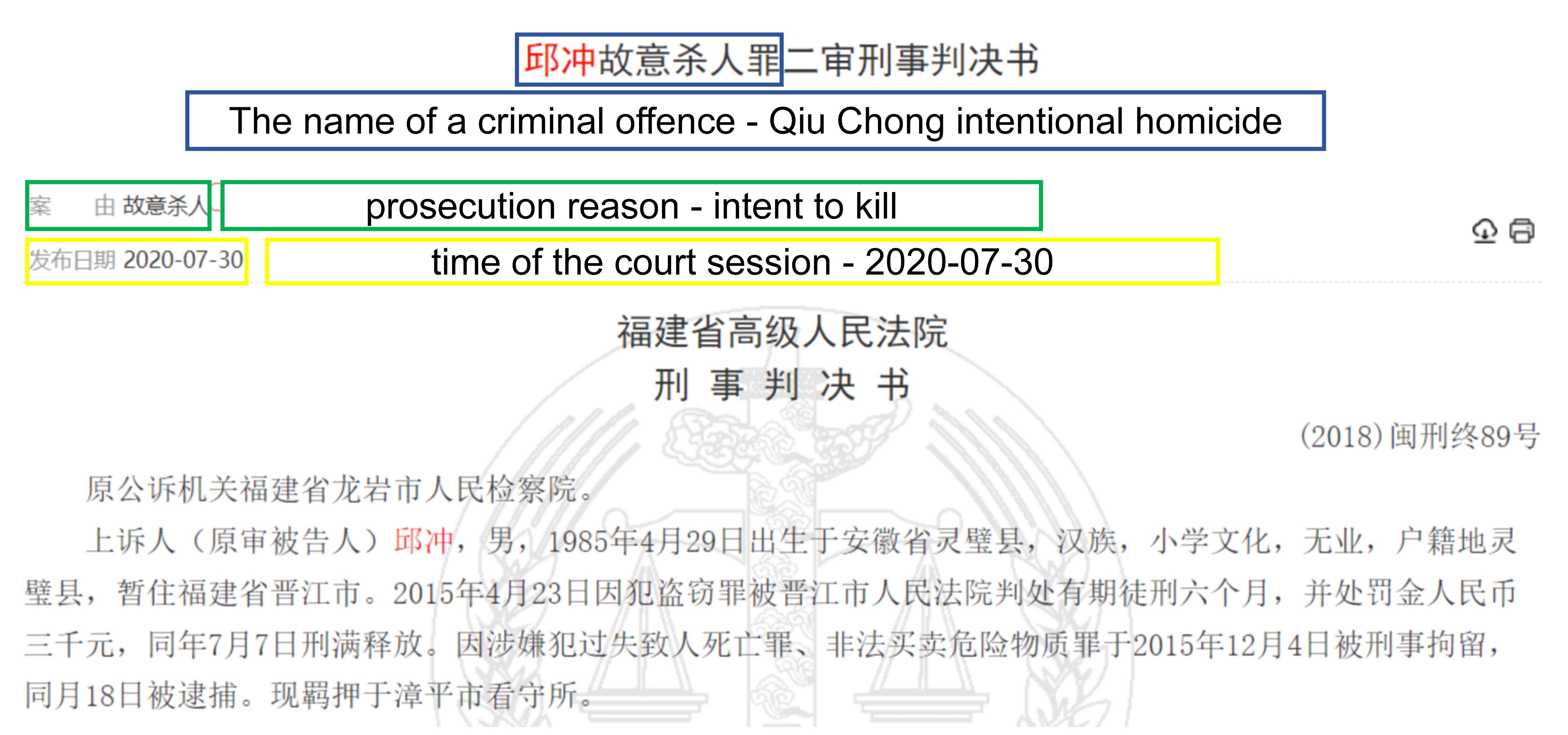}
\vspace{-5mm}   
\caption{The name of a case, time of the court session, prosecution reason on the China Judgments Online.}
\label{fig:caipanwang}
\vspace{-2mm}
\end{figure}

The criminal case data analyzed in this study were obtained from the China Judgments Online \autocite{SPC2023} platform. Figure \ref{fig:caipanwang} presents one of the cases. In light of the fact that the Amendment (XI) to the Criminal Law came into force on March 1, 2021, all judicial decisions used herein pertain to cases adjudicated after this date. During this period, due to concerns over information security and privacy protection, the number of publicly available criminal case judgments was limited, and the publication policy is outside the control of the authors. As a consequence, certain law articles correspond to only a very small number of cases, while some provisions have no publicly accessible cases at all. To mitigate this constraint, we exhaustively collected \emph{all} publicly accessible criminal judgments under Amendment (XI) that matched our statutory subset, yielding a dataset of 1,170 case records covering 85 distinct specific-law articles. Among them, 997 case records were designated for the training set and 173 for the test set. The training set encompassed 49 distinct law articles, whereas the test set covered 85 law articles (zero-shot number = 36). Despite this moderate cardinality in absolute terms, the resulting law article recommendation task is already highly non-trivial: the label space is large and severely imbalanced, and, as shown in Section~4.4, several classical text-classification baselines tend to collapse to the majority label. Throughout all experiments, the retriever in our framework still operates over all 452 law-article nodes in CLAKG, so retrieving the correct article is not artificially simplified by a small candidate pool. The Adjudicated Cases Knowledge Graph was subsequently constructed based on the training data, following the approach outlined in Section~3.1.

\subsection{CLAKG Construction and Graph Embedding}
We employed the method described in Section 3.1 to integrate the LAKG and the ACKG, resulting in the CLAKG. The types and numbers of nodes as well as relationships in CLAKG are presented in Table \ref{table9}.
\begin{table}[h]
\caption{Entity and relation types with their Numbers in CLAKG.}
\label{table9}
\begin{center}
\begin{tabular}{p{0.35\textwidth}|p{0.15\textwidth}|p{0.25\textwidth}|p{0.1\textwidth}}
\textbf{Entity Type} & \textbf{Number} & \textbf{Relation Type} & \textbf{Number} \\ \hline
Original article of law & 452 & Key & 7854 \\ \hline
Key information & 3405 & Id & 452 \\ \hline
Law article id & 452 & Agree With & 4820 \\ \hline
The name of a case & 997 & Applicable Law & 4118 \\ \hline
Time of the court session & 997 & Occur in Time & 997 \\ \hline
Prosecution reason & 997 & Reason & 997 \\ \hline
The specifics of the case & 997 & Detail & 997 \\ 
\end{tabular}
\end{center}
\end{table}
\begin{figure}[h]
\vspace{-8mm}
 \centering
\includegraphics [%
    width=1\textwidth] {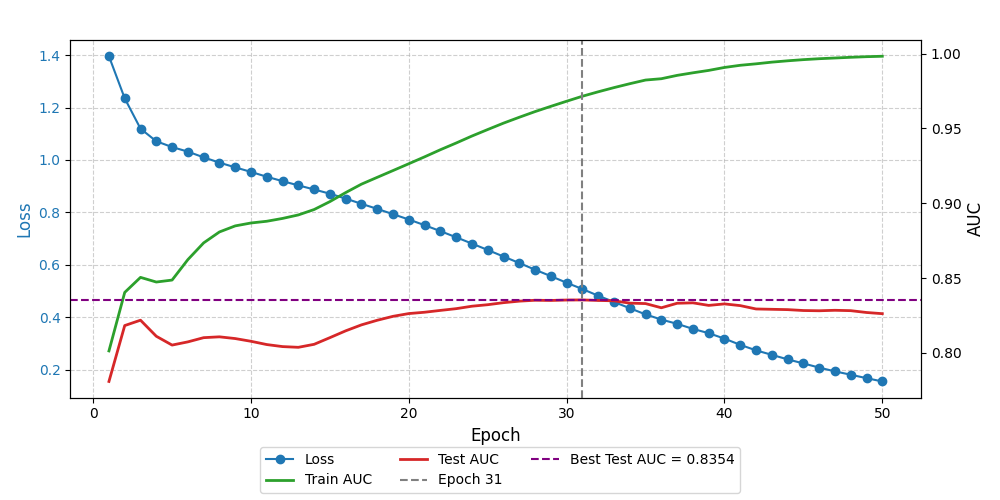}
\caption{Train loss and AUC over epochs.}
\label{fig:TrainandtestAUCoverepochs}
\vspace{-4mm}
\end{figure}

We employed the RGCN model described in Section \ref{rgcn} to perform graph embedding preprocessing on CLAKG. Our hyperparameter choices were as follows: $h_{dim} = 16$, $test\_size = 0.2$, $learning\_rate = 0.01$, $num\_epochs = 50$. We selected the model results corresponding to the highest Test AUC as the final graph embedding outcome. The Train AUC and Test AUC curves are presented in Figure \ref{fig:TrainandtestAUCoverepochs}. Upon examining this figure, we observed that  the Train AUC exhibited an overall upward trend. The Test AUC increased when $Epoch \leq 31$ and decreased when $Epoch \geq 31$. Consequently, we selected the model results at Epoch 31 as the final graph embedding outcome.\\
\subsection{Case Study}
We illustrate the law article recommendation process based on LLM and CLAKG using the case of "Zhang Yue's offenses of bribery and abuse of power" as an example. Initially, the user inputs the details of a new case:
\begin{quote}
\textit{During Zhang Yue's tenure as cultural administrator and grid member of Yongfeng Village, Zhang Yue manipulated residential information to secure illicit benefits, accepting 72,500 yuan in bribes. After the crime was uncovered, he voluntarily admitted his wrongdoing and returned the funds.}
\end{quote}
The system then retrieves keywords pertinent to the case using the methodology outlined in Section 3.2.2, identifying the following terms: \texttt{[`accepting bribes', `abuse of power', `bribery']}. Subsequently, the system locates the corresponding node IDs within the CLAKG and obtains the graph embedding vectors for these nodes as described in Section 4.2. Utilizing the historical case information and the candidate law articles retrieval algorithm specified in Section 3.2.3, the system identifies the candidate law articles and their associated cases.\\
The system consolidates the above information into a prompt using the method described in Section 3.2.4, and submits it to the LLM to carry out the law article recommendation task. Ultimately, the LLM outputs ``Article 385" as the recommended law article. The result produced by the LLM is consistent with the outcome recorded in the court judgment.

\subsection{Experimental Results}
\textcolor{blue}{In this subsection, we organize the baselines into three comparison groups to evaluate the proposed LLM+CLAKG framework on the law article recommendation task. Group (i) consists of classical text-classification models that directly map judgment texts to law-article labels (e.g., BERT, GRU, and DPCNN). Group (ii) includes LLM-based and generic RAG-style pipelines adapted from recent retrieval-augmented generation paradigms (e.g., plain LLM prompting, TFIDF-RAG, Graph-RAG, and Light-RAG), which represent the current mainstream approach for knowledge-intensive NLP tasks. Group (iii) presents an LLM-agnostic ablation across backbone models, where we instantiate the same CLAKG-RAG pipeline with two representative backbones—one mainstream proprietary model (GPT-4 \cite{openai2023chatgpt}) and one strong open-source model (Llama-3.1-405B-chat \cite{grattafiori2024llama3herd})—and evaluate each backbone with and without CLAKG-based augmentation. This three-group organization enables a clearer positioning of our method with respect to (a) traditional neural classifiers, (b) recent generic RAG baselines, and (c) the incremental benefit of CLAKG when plugged into different LLM backbones under a unified evaluation setting.}

\textcolor{blue}{Our primary objective here is to quantify the performance gain brought by our knowledge-graph--based RAG framework (CLAKG-RAG) for law article recommendation. In principle, any LLM can serve as a baseline for this task by directly prompting it with case facts and asking it to output applicable law articles. Given the rapid proliferation of LLM families and the non-trivial monetary and engineering cost of large-scale evaluations, a comprehensive comparison against ``all'' available LLMs is practically and economically infeasible. We therefore adopt a representative strategy: we keep the CLAKG construction, retrieval pipeline, and prompting scheme fixed, and instantiate them with a small but diverse set of competitive backbone models to demonstrate both effectiveness and generality. This design also reflects the model-agnostic nature of our framework: in principle, any sufficiently capable instruction-following LLM (open-source or proprietary) can be plugged into the same CLAKG-RAG pipeline in future work.}

\subsubsection{Comparison with classical text-classification baselines}

To analyze the impact of data scarcity and label imbalance on law article recommendation, we first compare our framework with three widely used neural models for text classification, namely BERT~\cite{devlin2019bert}, GRU~\cite{cho2014learning} and DPCNN~\cite{johnson2017deep}. All of them treat law article recommendation as a text classification problem on the raw judgment documents. In this setting, our method corresponds to the LLM-based recommendation framework built on CLAKG \emph{with} case information.

\begin{table}[h!]
\centering
\caption{Accuracy of classical models vs.\ our LLM+CLAKG.}
\begin{tabular}{l|c}
\hline
\textbf{Model} & \textbf{Accuracy} \\ \hline
BERT (baseline) & 0.289 \\ \hline
GRU (baseline) & 0.427 \\ \hline
DPCNN (baseline) & 0.495 \\ \hline
\textbf{LLM + CLAKG (ours, with case)} & \textbf{0.694} \\ \hline
\end{tabular}
\label{table:model_accuracy1}
\end{table}

As shown in Table~\ref{table:model_accuracy1}, the accuracy of the BERT baseline is only 0.289. A closer inspection of the predictions reveals that BERT tends to assign almost all test cases to the majority label ``Article~133''. This collapse is likely caused by the small size of the training set, the severe label imbalance, and the difficulty of capturing fine-grained legal distinctions from limited data.

The GRU baseline achieves an accuracy of 0.427. As a recurrent model designed for sequential data, GRU can capture some contextual information in the judgment texts, but it still struggles to distinguish between articles under highly imbalanced labels. The DPCNN baseline, a deep pyramid convolutional architecture tailored for text classification, further improves the accuracy to 0.495 by modeling hierarchical local patterns, yet it remains far from satisfactory. These observations indicate that, despite the moderate dataset size, the law article recommendation task on our corpus remains challenging: the label space is large (85 distinct articles in the test set) and highly imbalanced, and classical text-classification models tend to collapse to the majority label under such conditions.

In contrast, our LLM+CLAKG model (with case information) reaches an accuracy of 0.694, significantly outperforming all classical baselines. This suggests that (i) explicitly introducing legal structure via CLAKG and (ii) leveraging LLMs to reason over graph-grounded evidence can effectively mitigate the negative impact of insufficient data and label imbalance, whereas pure text-classification models are prone to majority-label bias on this dataset.

\textcolor{blue}{The classical text-classification models have the advantage of being conceptually simple, computationally efficient, and easy to deploy once a sufficiently large and well-balanced labeled dataset is available. However, they operate purely at the document-text level and do not encode the structured relations between cases, key legal elements, and law articles. As a result, they are highly sensitive to label imbalance and data sparsity, and tend to collapse to majority labels in our setting. In contrast, our LLM+CLAKG framework explicitly injects legal structure through the knowledge graph and leverages LLM reasoning over graph-grounded evidence. This design yields better robustness under limited and imbalanced data, at the cost of a more complex pipeline that requires KG construction and LLM inference.}

\subsubsection{Comparison with RAG-based baselines}

To investigate the impact of hallucinations in LLMs and to compare with more recent law article recommendation paradigms, we further evaluate several RAG-based methods: a plain LLM~\cite{openai2023chatgpt} (OpenAI's ChatGPT-4.0), TFIDF-RAG~\cite{aizawa2003information}, Graph-RAG~\cite{edge2024from} and Light-RAG~\cite{guo2024lightragsimplefastretrievalaugmented}. These methods instantiate retrieval-augmented generation pipelines that have recently attracted considerable attention, and here they are adapted to the law article recommendation setting. In this group of experiments, our method corresponds to the LLM-based recommendation framework built on CLAKG \emph{without} using case nodes (only statutory information), denoted as LLM+CLAKG (ours, no case).

\begin{table}[h]
\centering
\caption{Accuracy comparison with LLM and RAG-based baselines.}
\label{tab:model_accuracy_comparison2}
\begin{tabular}{l|c}
\hline
\textbf{Model} & \textbf{Accuracy} \\ \hline
LLM (baseline) & 0.549 \\ \hline
TFIDF-RAG (baseline) & 0.597 \\ \hline
Graph-RAG (baseline) & 0.179 \\ \hline
Light-RAG (baseline) & 0.636 \\ \hline
\textbf{LLM + CLAKG (ours, no case)} & \textbf{0.676} \\ \hline
\end{tabular}
\end{table}

As shown in Table~\ref{tab:model_accuracy_comparison2}, a plain LLM without external legal data already achieves an accuracy of 0.549, which is much higher than that of the classical text-classification models. This confirms that large pretrained language models are able to capture a considerable amount of legal knowledge from their pretraining corpora. However, the LLM still suffers from hallucinations, occasionally citing inapplicable or even nonexistent articles.

TFIDF-RAG improves the accuracy to 0.597 by retrieving judgment documents and law texts based on lexical similarity and feeding them as additional context to the LLM. Light-RAG further increases the accuracy to 0.636 by using a more efficient retrieval and organization strategy. Both methods demonstrate the benefit of grounding the LLM on retrieved legal documents. In contrast, Graph-RAG performs poorly on our dataset (0.179), which we attribute to its generic graph construction and retrieval strategy not aligning well with the fine-grained structure of Chinese criminal law and the limited size of our corpus.

Our LLM+CLAKG (ours, no case) achieves an accuracy of 0.676, outperforming the plain LLM, TFIDF-RAG, Graph-RAG and Light-RAG. Compared with TFIDF-RAG and Light-RAG, which mainly rely on surface-level text similarity, our approach matches new cases to law articles via \emph{key information nodes} in CLAKG, and uses graph embeddings (Section \ref{rgcn}) to exploit the topological structure information of the graph. As a result, candidate law articles are selected based on their structured legal relevance rather than purely lexical overlap.

Moreover, by constraining the LLM to reason over candidate articles and historical cases retrieved from CLAKG (Section \ref{subsec324}), the system is grounded in verifiable law articles and case information, which mitigates hallucinations in article citations and explanations more effectively than generic RAG pipelines. In summary, the comparison with both classical classifiers and recent RAG-based methods shows that our LLM+CLAKG framework better captures the underlying legal reasoning and achieves consistently higher recommendation accuracy.

\textcolor{blue}{RAG-based methods mitigate hallucinations by retrieving external texts and are relatively easy to adapt across domains. TFIDF-RAG and Light-RAG work well when lexical similarity is a good proxy for relevance, but they mainly exploit surface-level word overlap and do not explicitly model legal entities or relations; their performance may degrade when key legal distinctions are not reflected in simple bag-of-words similarity. Graph-RAG introduces a graph structure, yet its generic document graph and retrieval strategy are not tailored to the fine-grained structure of Chinese criminal law in our corpus, which limits its effectiveness. In contrast, our proposed LLM+CLAKG framework constructs a task-specific CLAKG schema centered on law article recommendation. This schema leverages legally meaningful key information in law articles together with adjudicated case information to precisely characterize the relations among different law articles as well as the links between cases and law articles. During the retrieval-augmentation stage, this semantic and structured information is injected into the LLM through prompt engineering. In this way, we substantially mitigate hallucinations of large language models and significantly improve the accuracy of law article recommendation. The trade-off is that our approach requires schema design and KG construction effort, as well as maintaining the KG over time, but it delivers consistently higher accuracy of legal reasoning than these more generic RAG pipelines.}

\subsubsection{Ablation Study}
In this ablation study, we evaluate the impact of incorporating different types of information on the performance of LLMs for law article recommendation tasks. Specifically, we compare the performance of the baseline models with and without additional case and law article information.

We conducted experiments using both the proprietary GPT-4 model and the open-source Llama-3.1-405b-chat model. The baseline results for GPT-4 showed an accuracy of 0.549, while incorporating case-specific information with CLAKG (Case Law and Statutory Knowledge Graph) improved the accuracy to 0.694. Similarly, for the Llama-3.1-405b-chat model, the baseline performance was 0.619, and adding reference case and statutory information boosted the accuracy to 0.786.

These results clearly demonstrate that the inclusion of relevant information, such as reference cases, significantly enhances model performance across both proprietary and open-source models. This validates the robustness and versatility of our approach, showing its applicability not only to high-performing proprietary models but also to widely available open-source models.
The detailed results are summarized in Table \ref{tab:ablations}.
\begin{table}[ht]
\centering
\caption{Results of different models in the ablation study.}
\begin{tabular}{l|c}
\textbf{Model} & \textbf{Accuracy} \\
\hline
LLM (GPT-4, baseline) & 0.549 \\ \hline
\textbf{LLM + CLAKG (GPT-4, with case)} & \textbf{0.694} \\ \hline
LLM (llama-3.1-405b-chat, baseline) & 0.619 \\ \hline
\textbf{LLM + CLAKG (llama-3.1-405b-chat, with case)} & \textbf{0.786} \\
\end{tabular}
\label{tab:ablations}
\end{table}\\
This ablation study highlights the significant impact that additional legal context, such as case and law article information, can have on the performance of LLMs for law article recommendation tasks, confirming the effectiveness of our proposed methodology.\\

\section{Conclusion}
\label{sec:conclusion}
\textcolor{blue}{This paper presents an efficient law article recommendation approach that combines a Case-Enhanced Law Article Knowledge Graph (CLAKG) with a Large Language Model (LLM). CLAKG integrates law articles and adjudicated cases under a task-oriented schema, with a rich topological structure formed by usage relations and shared key information extracted from judgments. By grounding the LLM in this graph-structured macro semantics, the proposed framework produces more accurate recommendations of law articles and related cases, while mitigating typical LLM hallucinations. Extensive experiments against strong text-classification baselines (BERT, GRU, DPCNN) and recent RAG-based methods (TFIDF-RAG, Graph-RAG, Light-RAG) demonstrate that our approach effectively alleviates issues such as insufficient data, label imbalance, and hallucination in LLMs, leading to substantial gains in recommendation accuracy.}

\textcolor{blue}{Looking ahead, we plan to expand CLAKG in both scale and scope. As more criminal judgments under Amendment (XI) become publicly available, we will be able to evaluate the framework on larger and more diverse corpora. When appropriate data access is possible, extending the evaluation to other jurisdictions will further test the scalability and generality of our approach. Beyond passively waiting for new real-world judgments, a complementary research direction is to explore carefully controlled data augmentation via large language models: LLMs can be used to generate hypothetical criminal case fact patterns conditioned on specific law articles, after which legal experts verify, correct, or discard these synthetic cases before they are integrated into CLAKG as new training and evaluation samples. Such human-validated synthetic cases have the potential to mitigate label imbalance and sparsity while preserving legal plausibility.}

\textcolor{blue}{It is important to emphasize that the closed-loop human--machine collaboration proposed in this paper is a deployment-level mechanism for maintaining and improving CLAKG over time, rather than a separate algorithmic baseline. Accordingly, our quantitative experiments in Section~\ref{sec:experiment} focus on recommendation accuracy under different model variants, assuming this closed-loop mechanism is in place. A natural next step is to design quantitative evaluation protocols for the human feedback loop itself---for example, tracking the reduction in the proportion of recommendations that require correction, measuring expert time and effort, and analyzing how the evolving CLAKG affects downstream accuracy over longer periods.}

\textcolor{blue}{The current CLAKG construction still relies on LLM-based extraction of key information and case descriptions. Although we constrain the LLM with a predefined schema and strict post-processing, extraction errors inevitably remain and primarily manifest as missing or noisy links in the graph. Developing more robust and formally guaranteed information-extraction pipelines (e.g., combining supervised or rule-based extraction with LLMs) is therefore an important direction for future work. }

\textcolor{blue}{While our experiments focus on Chinese criminal law, a civil-law system with codified statutes, the proposed CLAKG--LLM framework is conceptually applicable to other legal systems as well. Adapting it to different jurisdictions, however, raises several challenges. First, the underlying legal structure may differ substantially: common-law systems rely more heavily on precedents and citation networks between cases, which would require extending the CLAKG schema with additional entity and relation types (e.g., precedent cases, holdings, ratios, and explicit citation links) and redesigning task-specific key information nodes. Second, legislative drafting conventions and document structures vary across countries (e.g., hierarchical numbering, cross-references, annotations), so the extraction--transformation--loading (ETL) pipeline and schema-design steps must be tailored to each legal corpus, as illustrated by recent work on Italian legislation~\cite{sovrano2020legal,colombo2025llm}. Third, applying the framework to other languages calls for high-quality LLMs and tokenization tools for the target language, as well as careful adaptation of prompts and extraction templates to local legal terminology. Despite these challenges, the overall architecture---schema-driven KG construction, graph-embedding-based retrieval, and KG-grounded LLM reasoning---remains unchanged, suggesting that our approach can serve as a blueprint for building similar systems in other jurisdictions.}

\section*{\textcolor{blue}{Data Availability}}

\textcolor{blue}{All source code, the constructed Case-Enhanced Law Article Knowledge Graph (CLAKG), and the processed law article and case datasets used in this paper are publicly available in a GitHub repository maintained by the authors:}

\begin{center}
\textcolor{blue}{\url{https://github.com/Songan-Lab/Law_Rec_CLAKG_LLM}}
\end{center}

\textcolor{blue}{The repository includes: (i) dump files of CLAKG, (ii) the de-identified structured data used in our experiments, and (iii) the full implementation of the proposed LLM--CLAKG law article recommendation framework together with instructions for reproducing the experimental results. The original raw judgments remain accessible through the official ``China Judgments Online'' website, subject to its access and privacy policies.}

\printbibliography

\end{document}